\documentclass[final,3p,times]{elsarticle}

\usepackage{amssymb}
\usepackage{amsmath}
\usepackage{multirow}

\usepackage{xcolor}

\usepackage[normalem]{ulem}

\usepackage{url}

\journal{Physics of the Dark Universe}

\begin{document}

\begin{frontmatter}

\title{Power Spectrum Emulators from Neural Networks and Tree-Based Methods} 

\author[L1,L2]{Andrei Lazanu} 
\affiliation[L1]{organization={Department of Physics and Astronomy, University of Manchester},
            city={Manchester},
            postcode={M13 9PL}, 
            country={United Kingdom}}

\affiliation[L2]{organization={Institute of Space Science - INFLPR Subsidiary},
            city={Magurele},
            postcode={077125}, 
            country={Romania}}

\begin{abstract}
We use two subsets of 2000 and 1000 \textsc{Quijote} simulations to build two power spectrum emulators, allowing for fast computations of the non-linear matter power spectrum. The first emulator is built in terms of seven cosmological parameters: the matter and baryon fraction of the energy density of the Universe $\Omega_m$ and $\Omega_b$, the reduced Hubble constant $h$, the scalar spectral index $n_s$, the amplitude of matter density fluctuations $\sigma_8$, the total neutrino mass $M_{\nu}$ and the dark energy equation of state parameter $w$, on scales $k \in [0.015,1.8]\,h/ \rm{Mpc^{-1}}$. The power spectra can be directly determined at redshifts 0, 0.5, 1, 2 and 3, while for intermediate redshifts these can be interpolated. The second emulator is based on five cosmological parameters, $\Omega_m$, $h$, $n_s$, $\sigma_8$ and the amplitude of equilateral non-Gaussianity $f_{\rm NL}^{\rm eq}$, at redshifts 0, 0.503, 0.733, 0.997 for $k \in [0.015,1.8]\,h/ \rm{Mpc^{-1}}$. The emulators are built on machine learning techniques. In both cases we have investigated both neural networks and tree-based methods and we have shown that the best accuracy is obtained for a neural network with two hidden layers. Both emulators achieve a root-mean-squared relative error of less then 5\% for all the redshifts considered on the scales discussed.
\end{abstract}


\end{frontmatter}

\section{Introduction}

Cosmology has become in the last few decades a precision science, mainly due to the unprecedented development of instruments capable of making cosmological measurements with exquisite accuracy. In particular, the $\Lambda$CDM model \cite{RevModPhys.61.1}, a six-parameter model based on the spontaneous generation of primordial quantum fluctuations followed by an early-time inflationary epoch has been shown to accurately describe the cosmological observations from the Cosmic Microwave Background (CMB) data provided by COBE, WMAP \cite{2003ApJ...583....1B}, \textit{Planck} \cite{Akrami:2018vks},  the Atacama Cosmology Telescope \cite{ACT:2020gnv,ACT:2020frw,ACT:2023kun,ACT:2023dou} and the South Pole Telescope \cite{SPT-3G:2022hvq,SPT:2023jql}. These have placed increasingly stringent constraints on the parameters of this model, as well as on its extensions. These constraints will become even tighter in the future, with observations from CMB-S4 \cite{CMB-S4:2016ple} and the Simons Observatory \cite{SimonsObservatory:2018koc}.  

In parallel, the late-time matter and galaxy distributions, the large-scale-structure (LSS), can provide complementary information to the CMB, with valuable information coming from its three-dimensional nature given by the redshift, in addition to the distribution of galaxies on the sky. Current LSS probes, such as the Dark Energy Survey \cite{DES:2021wwk}, the Kilo-Degree Survey \cite{Heymans:2020gsg}, as well as experiments in progress, and future ones, such as Euclid \cite{Laureijs:2011gra}, the Vera C. Rubin Observatory Legacy Survey of Space and Time \cite{Abell:2009aa}, the Square Kilometre Array \cite{Jarvis:2015tqa}, the Nancy Grace Roman Space Telescope \cite{Eifler:2020vvg} will be able to improve our understanding even further. The exquisite accuracy in the measurements of these experiments must be matched by the theoretical modelling. 

However, modelling of non-linear scales remains a challenge: perturbation theories break down at a certain scale, and one has to rely on phenomenological models, fit on simulations or work directly with the desired simulations. In recent years, a large number of simulations have been produced, and their generation and processing has benefited from important progress in machine learning usage. Machine learning has been used in a large number of fields in cosmology, including CMB \cite{Caldeira:2018ojb, Chanda:2021qbf}, LSS \cite{Rodriguez:2018mjb,Lucie-Smith:2020ris,Lin:2021dim,Xu:2021pdp}, reionization and 21cm \cite{Shimabukuro:2017jdh,Huang:2020wvt,Floss:2023ylq,Saxena:2024rhu}, gravitational lensing: weak lensing \cite{Ribli:2018kwb,ZorrillaMatilla:2020doz}, strong lensing \cite{Jacobs:2017xhn,Park:2020eat}, redshift prediction \cite{Collister:2003cz,Eriksen:2020diu}, parameter estimation \cite{Alsing:2019xrx, Kostic:2021tyw, Lazanu:2021tdl} etc.

In spite of the significant advances in computational power, these remain expensive to run and phenomenological models rely on specific simulations; new simulations, or simulations taking into account additional parameters can require a complete rethink of the model. Hence, emulators have been built to mitigate these issues. These are particularly useful where a specific cosmological quantity (such as the power spectrum in this work) needs to be repeatedly calculated, for example, in the case of placing parameter constraints using Markov-Chain-Monte-Carlo (MCMC) methods. The techniques employed to build such quantities are varied and have started with the pioneering fitting formulae, such as HALOFIT and its extensions \cite{Smith:2002dz, Takahashi:2012em,Mead:2020vgs}, work involving Bayesian likelihood estimators, linear algebra data reduction techniques (such as Principal component analysis) and MCMC methods \cite{Habib:2007ca,Heitmann:2006hr,Heitmann:2008eq,Heitmann:2009cu,Lawrence:2009uk, Heitmann:2013bra,Euclid:2020rfv}, to the use of Gaussian processes \cite{Arnold:2021xtm,Moran:2022iwe,Saez-Casares:2023olw,Ho:2023alj,Gunther:2025xrq} and neural networks methods \cite{Agarwal:2012ew,Agarwal:2013aea,Ramanah:2020vyl,Angulo:2020vky,Ramachandra:2020lue, Kaushal:2021hqv, Jamieson:2022lqc, Bolliet:2023sst,Piras:2023aub, Bonici:2023xjk,Mauland:2023pjt,Fiorini:2023fjl,Jense:2024llt,Trusov:2024mmw,Bai:2024cgt,Kaushal:2025gwe} and a combinations of analytical and numerical fits to the data \cite{Mohammed:2014lja, Winther:2019mus, Bose:2022vwi,Bartlett:2023cyr,Sui:2024wob}.
These methods rely on a large number of simulations and the power spectra involve both $\Lambda$CDM cosmology and its extensions, such as modified gravity or massive neutrinos.

In this paper, we show how to build two emulators for the non-linear matter power spectrum, by using machine learning techniques starting from $N$-body simulations. The first one is based on simulations where seven cosmological parameters are varied: the dark matter density parameter $\Omega_m$, the baryon density parameter $\Omega_b$, the reduced Hubble constant $h$, the scalar spectral index $n_s$, the amplitude of the root-mean-square matter fluctuation averaged over a sphere of radius $8h^{-1} \mathrm{Mpc}$ $\sigma_8$, the sum of the neutrino masses $M_\nu$ and the equation of state parameter for dark energy $w$.

The second emulator varies five cosmological parameters,  $\Omega_m$,  $h$,  $n_s$, $\sigma_8$ and the amplitude of equilateral primordial non-Gaussianity $f_{\rm NL}^{\rm eq}$. $\Omega_b=0.049$ is fixed for these runs.

The paper is structured as follows: in Section \ref{sec:models} we briefly describe the two extensions to $\Lambda$CDM considered in this work, in Section \ref{sec:sims} we present the simulations used in this work and the machine learning models employed in the analysis;  Section \ref{sec:results_mnu} presents our results from the fittings and the performances of each of the seven-parameter emulator which involves neutrino masses and the dark energy equation of state parameter, and  Section \ref{sec:results_fnl} is devoted to the results for the emulator involving equilateral primordial non-Gaussianity; in Section 6 we investigate the performance of the emulator by performing a MCMC analysis and comparing it to \texttt{HALOFIT}; we conclude in Section \ref{sec:conc}. 

\section{$\Lambda$CDM extensions}
\label{sec:models}
\subsection{Massive neutrinos and dark energy equation of state parameter}

Neutrinos  appear as a triplet and the oscillation experiments have confirmed that at least two types have non-zero masses \cite{ParticleDataGroup:2024cfk}. Cosmological probes can constrain the sum of the neutrino masses -- CMB constraints yield \cite{Planck:2018vyg}
$M_\nu=\Sigma m_\nu \lesssim 0.12 \, \text{eV}$ at 95\% confidence level, while LSS can constrain this further, to \cite{DESI:2025ejh} 
$M_\nu \lesssim 0.0642 \, \text{eV}$.

The late-time cosmic acceleration, which has been experimentally observed \cite{SupernovaSearchTeam:1998fmf} can be accounted for by the presence of dark energy. Its equation of state parameter, $w$, defined as the ratio between the pressure and the density of dark energy, quantifies how the Universe is expanding. The current constraints from the CMB  \cite{Planck:2018vyg} and LSS  \cite{DESI:2024mwx} are given by $w=-1.028 \pm 0.031$ and  $w =-0.99^{+0.15}_{-0.13}$ respectively. Both these measurements are compatible with $w=-1$, which corresponds to the cosmological constant. Hence, values $w \ne -1$, could point out to new physics beyond $\Lambda$CDM.

\subsection{Equilateral primordial non-Gaussianity}
Although the $\Lambda$CDM model is based on Gaussian scale-invariant primordial quantum fluctuations, most inflationary models predict some level of non-Gaussianity, which appears through a three-point function (bispectrum) correction at the primordial level. Indeed, for purely Gaussian fields, the two-point correlation function contains all the relevant information, and the bispectrum vanishes. The non-vanishing primordial bispectrum modifies the spectra throughout the history of the Universe, and in particular it has a contribution at the power spectrum level. As the bispectrum is three-dimensional (defined in Fourier space by the magnitudes of three wavevectors forming a triangle), its shape and amplitude ($f_{\rm NL}$) both play important roles in the behaviour of cosmological quantities. Based on the details of the inflationary model considered, it peaks at various triangular configurations (see \cite{Chen:2006nt} for a review). These shapes, the most popular being the local, equilateral and orthogonal, have been tightly constrained by current CMB data from \textit{Planck}  \cite{Jung:2025nss}, and in the future LSS surveys have the potential to tighten the constraints even further \cite{Meerburg:2016zdz,Karagiannis:2018jdt}. Their effects can be measured through the power spectrum and higher-order correlation functions.

In this paper, we focus on the equilateral type of non-Gaussianity; the primordial bispectrum $B_{\Phi}^{\text{eq}}$  can be expressed in terms of the primordial power spectrum $P_{\Phi}$ as
\begin{align}
B_{\Phi}^{\text{eq}}&(k_1,k_2,k_3)=6 f^{\rm eq}_{\rm NL} \left\{-[P_{\Phi}(k_1)P_{\Phi}(k_2)+\text{2 perms} ]\right. \nonumber \\
&-2[P_{\Phi}(k_1)P_{\Phi}(k_2)P_{\Phi}(k_3)]^{2/3} 
+[P_{\Phi}^{1/3}(k_1)P_{\Phi}^{2/3}(k_2)P_{\Phi}(k_3)+\text{5 perms}]\left. \right\} \,,
\end{align} 
where the parameter $f_{\rm NL}^{\rm eq}$ represents the amplitude of primordial non-Gaussianity.
Cosmological probes are measuring this quantity, with current constraints given by $f^{\rm eq}_{\rm NL} = 6 \pm 46$. The detection of a non-zero equilateral-type of non-Gaussianity would point out to specific inflationary models, and in particular to single-field inflation with non-standard kinetic terms.

\section{Numerical simulations and methodology}
\label{sec:sims}

We built the emulators using a subset of the  \textsc{Quijote} simulations \cite{Quijote_sims}, a public suite  of 88000 full $N$-body simulations, specifically run to quantify the information content on cosmological observables and to provide enough statistics to train machine learning algorithms. They have been run 
using the TreePM code Gadget-III \cite{Springel:2005mi} in boxes of sides of $1\, {\rm Gpc}/h$. 

For the first emulator,  we made use of a subset of 2000 simulations where seven cosmological parameters are varied, namely $\Omega_m$, $\Omega_b$, $h$, $n_s$, $\sigma_8$, $M_\nu$ and $w$. They have been chosen using a latin-hypercube sampling method, a statistical method built for generating a near-random sample of parameter values from a multidimensional distribution \cite{10.2307/1268522}. The parameter ranges are given by: $\Omega_m \in [0.1,0.5]$, $\Omega_b \in [0.03,0.07]$, $h \in [0.5,0.9]$, $n_s \in [0.8,1.2]$, $\sigma_8 \in [0.6,1]$, $M_{\nu} \in [0.01, 1.0]\, \rm{eV}$ and $w \in [-1.3,-0.7]$. The input power spectrum and transfer functions are set at $z=127$ and have been obtained by rescaling the corresponding quantities derived at $z=0$ from CAMB \cite{Lewis:1999bs} using a method from Ref. \cite{Zennaro:2016nqo}, while the initial conditions have been generated using the Zel’dovich approximation.

The simulations have a volume of $1\, (h^{-1} {\rm Gpc})^3$ and contain $512^3$ dark matter particles. Outputs are provided at redshifts  0, 0.5, 1, 2, 3 and in this study we use all of them. Several cosmological quantities are calculated from the simulations, but in this work we only focus on the matter power spectra.

The \textsc{Quijote} simulations involving primordial non-Gaussianity \cite{Quijote-PNG,2022ApJ...940...71J}, are represented by a suite of 1000 simulations, where five parameters are sampled from a latin-hypercube, with $\Omega_m \in [0.1,0.5]$, $h \in [0.5,0.9]$, $n_s \in [0.8,1.2]$, $\sigma_8 \in [0.6,1]$ and $f^{\rm eq}_{\rm NL} \in [-600,600]$. The other cosmological parameters are fixed to $\Omega_b = 0.049$, $M_{\nu}=0$ and $w=-1$. For these,  the initial conditions for displacements and peculiar velocities are generated using second order perturbation theory, which in turn are used to assign positions and velocities to particles that are initially laid on a regular grid with the 2LPT code \cite{Scoccimarro:1997gr, Crocce:2006ve}. In this case, snapshots are provided at four redshifts: 0, 0.503, 0.733, 0.997. We note that one of the fundamental parameters of the $\Lambda$CDM model is $\omega_b \equiv \Omega_b h^2$, which has been tightly constrained by Big Bang nucleosynthesis. Therefore, as $\Omega_b$ is kept fixed in this emulator, the values of $\omega_b$ might become unphysical, limiting the practical applicability of this emulator.

We investigated different learning models for building our emulator, including deep neural networks and tree-based methods. We also checked whether the performance of the models can be improved by reducing the target size with data-reduction techniques, such as principal component analysis (PCA) \cite{Pearson01111901}. PCA consists of linearly transforming the data into a new orthogonal coordinate system such that the directions (principal components) that capture the largest variation in the data are ranked from highest to lowest. By keeping only a limited number of principal components, one can then reconstruct an approximation of the initial data matrix, at the same time keeping a controlled amount of the total variance of the data. 

We also considered several tree-based methods to solve this regression problem, in particular Random Forests, and the gradient boosting techniques CatBoost \cite{CatBoost}, XGBoost \cite{10.1145/2939672.2939785} and LightGBM \cite{NIPS2017_6449f44a} and we compared the results of these methods with the ones of the neural networks.

Apart from the overall root mean squared error determined by the optimizers, in order to see how the models perform at each scale, we have computed the root-mean-squared-relative error (RMSRE) for the test set for each value \textit{k}, defined as
\begin{equation}
{\rm RMSRE}(k)=\sqrt{\frac{1}{n_{\rm test}}\sum_{i \in \rm test\; set}\left(\frac{P_{\rm emulator}(k; \theta_i)}{P_{\rm simulation}(k; \theta_i)}-1\right)^2} \,,
\end{equation}
for each of the models considered, where $\theta_i$ represents the parameter vector of the ith power spectrum of the test set and $n_{\rm test}$ represents the size of the test set. For a good modelling of the simulations by our emulator, this quantity must be as close to 0 as possible for all the scales considered.

In the case of the neural networks, a 5-fold cross validation has been used to improve the robustness of the model. We note that for the PCA case, in order to avoid data leakage, the PCA reduction must be done after the data split, with the projection directions derived from the training set. For the tree-based models, we have additionally performed randomized cross-validation for some of the hyperparameters, again taking care of possible leakages in the pipeline.

\section{Results for neutrinos and dark energy equation of state parameter}
\label{sec:results_mnu}

In order to build the emulator, we trained learning models on the input data, consisting of the seven cosmological parameters varied in the simulations, while the output (target) variables are given by the corresponding power spectra, at fixed values of the wavevector $k \in[0.008900, 2.785996]$ $h/{\rm Mpc}$. As at large scales errors are expected to be large due to cosmic variance and at low scales the accuracy of the simulations is limited due to the resolution considered, we restrict the $k$-range to $k \in[0.015, 1.80]\, h/{\rm Mpc}$ for the training and predictions. Due to the limited redshift values available from the simulations, we chose not to use $z$ as an additional feature of the model, and we build separate models for each redshift showing how one can potentially interpolate between them.

The inputs are linearly scaled so that they are all in the interval $[0,1]$.  The outputs are represented by the matter power spectra, evaluated at fixed values of $k$. As the power spectra span several orders of magnitude at each redshift, $P(k) \in (10^2, 10^5)$ at $z=0$ to $(1, 10^4)$ at $z=3$, we apply logarithms in order to make all values $\mathcal{O}(1)$.
We started by randomly splitting the 2000 simulations into a training set and a test set. In order to improve the performance of the model by having a large enough training set, we only separated 100 simulations for testing purposes. For the methods considered, we have implemented two scenarios: one using the full output for the scales considered, in 285 points ($k$-values), and a compressed version of 11 points using PCA. In this case, more than 99.5\% of the variance of the matrix of power spectra is preserved.

For neural networks, several architectures have been investigated, and the best performing one consists of two hidden layers. For the full power spectrum output, the input vector consists of the seven features; there are two hidden layers with 2048 neurons each, with ReLU activation functions, followed by Dropout layers with rates of 15\% (15\% of the nodes are randomly set to 0 at each iteration), while for the 285 targets a linear activation layer has been applied. A mean-squared-error loss function and an Adam optimizer with a rate of $10^{-4}$ have been used. In order to perform the optimisation, and to split the data into training and validation sets, a K-fold splitting has been introduced with $k=5$. This technique increases the predictive power of the model, by splitting the training set into $k$ folds, each of which is used as a validation set and the model is trained $k$ times. Thus, $k=5$ models were run, and their predictions on the test set were averaged, yielding the final answer. 3000 training steps were set for each of the runs, but early stopping was also implemented, watching the decrease of the validation loss. Both the Dropout layers and early stopping were implemented to avoid overfitting.

A similar network architecture has been considered for the PCA output, with the same hidden layer sizes, but the Dropout rate has been set to 60\%. We note that here, in order to avoid data leakage, the PCA transformation matrix has been obtained from the training set in each of the k-folds, and the same transformation has been used on the validation and test sets too. Finally, the reconstruction of the  power spectrum was performed using the inverse transformation from each of the k-folds.  

\begin{figure}[!h]
\centering
\includegraphics[width=0.75\linewidth]{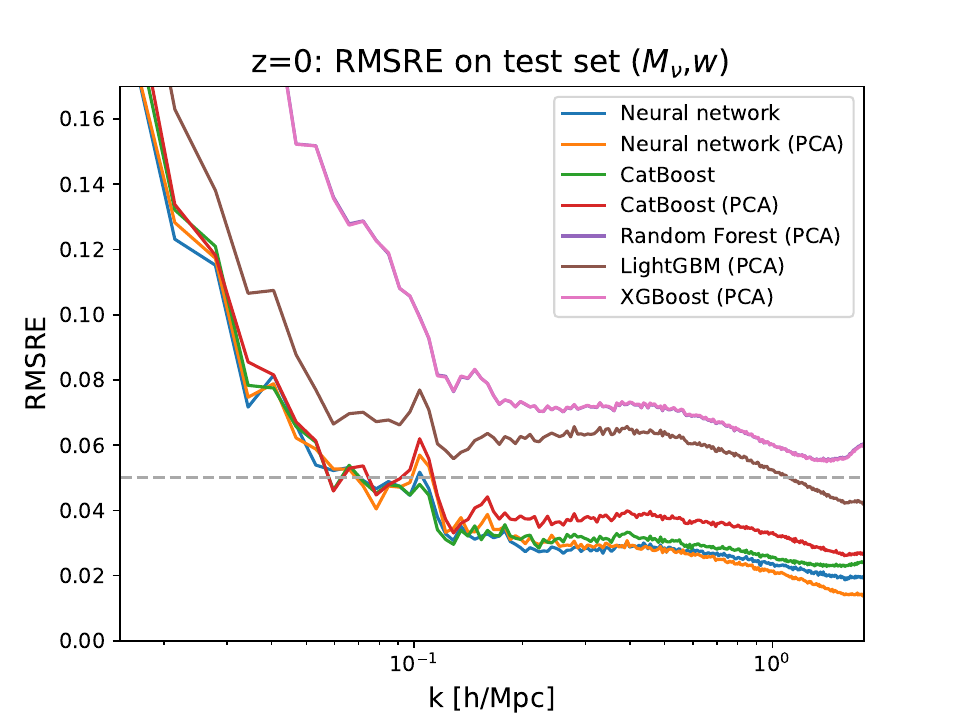} 
\caption{Comparison of RMSRE as a function of $k$ on the test set for the seven-parameter emulator ($\Omega_m$, $\Omega_b$, $h$,  $n_s$, $\sigma_8$,  $M_\nu$ and $w$) for the different learning models considered at $z=0$.}
\label{fig:nucomp}
\end{figure}

The overall training time of each of these two networks was around 15 minutes on a NVIDIA Tesla T4 GPU with 40 cores.

In the case of tree-based methods, we have considered the models described in Section \ref{sec:sims}, with the RMSRE plotted in Fig. \ref{fig:nucomp}. We noticed that, despite our best efforts, XGBoost and LightGBM did not produce satisfactory results, and hence we only worked with the neural networks and with CatBoost for the other redshifts. Moreover, modifying the CatBoost parameters did not seem to improve the performance of the model, with the exception of increasing the number of iterations, which also peaks around the value considered. The full CatBoost model ran for around 30 minutes with 5000 iterations on a single core, whereas for the PCA version the runtime is around 5 minutes with the same number of iterations.

The figure shows that the models involving the neural network are the best performing, with the PCA having a slightly better score. This is most likely due to the fact that the network has a significantly easier job of training on a reduced output. The RMSRE for these models is below 3\% for $k>0.15 h/{\rm Mpc}$ models and below 5\% for $0.10 h/{\rm Mpc} < k<0.15 h/{\rm Mpc}$. The performance is decreasing significantly for lower values of $k$, mainly due to the noise in the simulations. At these scales,  cosmic variance is important, and the sample size is too low to compensate for it. CatBoost performs slightly worse, with a decrease of the RMSRE less than 1\% to the neural networks for the scales of interest.
The overall RMSEs for the 4 best models for the redshifts considered are presented in Table \ref{tab:rmsenu}. These values are calculated on the power spectra (after applying the inverse PCA transform and exponentiating), which explains the larger values at lower redshifts.

\begin{table}[!h]
\begin{center}
\begin{tabular}%
{c|c|c|c|c|c}\hline
    Network & $z=0$ & z=0.5 & $z=1$ & $z=2$ & $z=3$ \\ \hline\hline
    Neural network &430.8 & 295.5 & 210.1 & 95.7 & 56.9  \\
    \hline
    Neural network (PCA) &358.4& 219.9 & 166.6 & 84.7 & 47.0 \\
    \hline
    CatBoost &478.2& 408.6 & 289.0 & 140.1 & 76.8 \\
    \hline
    CatBoost (PCA) &605.7& 396.9 & 282.6 & 137.8 & 75.3  \\
    \hline
\end{tabular}
\caption{The RMSE for the 4 best models considered at the redshifts of the simulations (0, 0.5, 1, 2 and 3) for the emulator with seven cosmological parameters: $\Omega_m$, $\Omega_b$, $h$,  $n_s$, $\sigma_8$,  $M_\nu$ and $w$.}
\label{tab:rmsenu}
\end{center}
\end{table}

\begin{figure}[!h]
\centering
\includegraphics[width=0.75\linewidth]{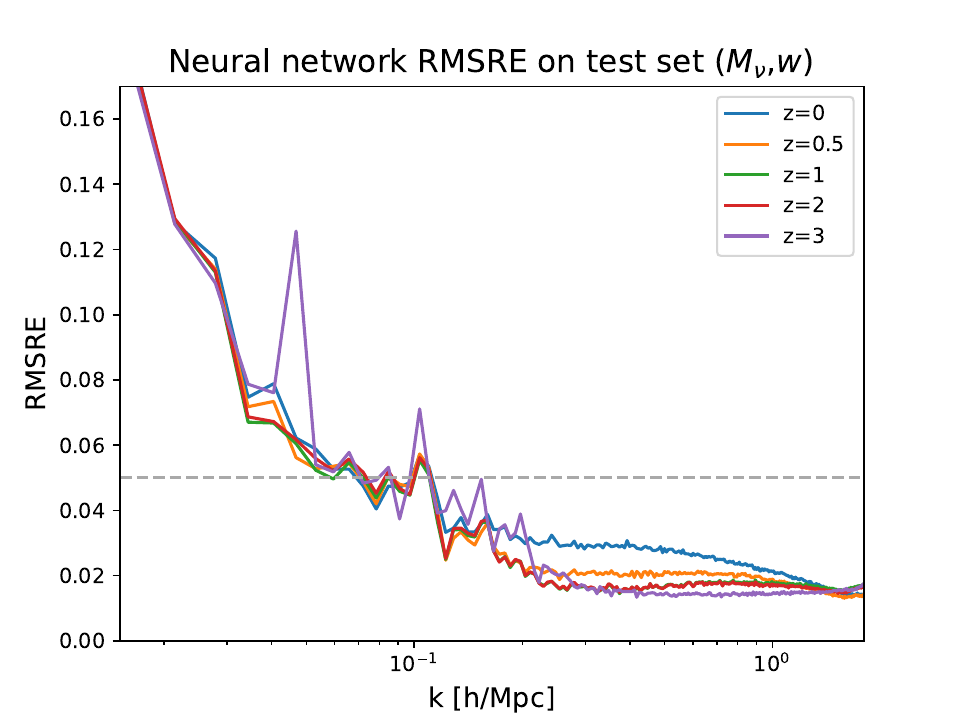} 
\caption{RMSRE for the neural network (with PCA) for redshifts 0, 0.5, 1, 2 and 3 for the emulator involving the sum of neutrino masses and the dark energy equation of state}.
\label{fig:nu_nn}
\end{figure}

As the neural network involving the PCA with 11 principal components systematically produces the best RMSRE, we only plot that for clarity at all redshifts, with the results shown in Fig. \ref{fig:nu_nn}.

\begin{figure}[!h]
\centering
\includegraphics[width=0.99\linewidth]{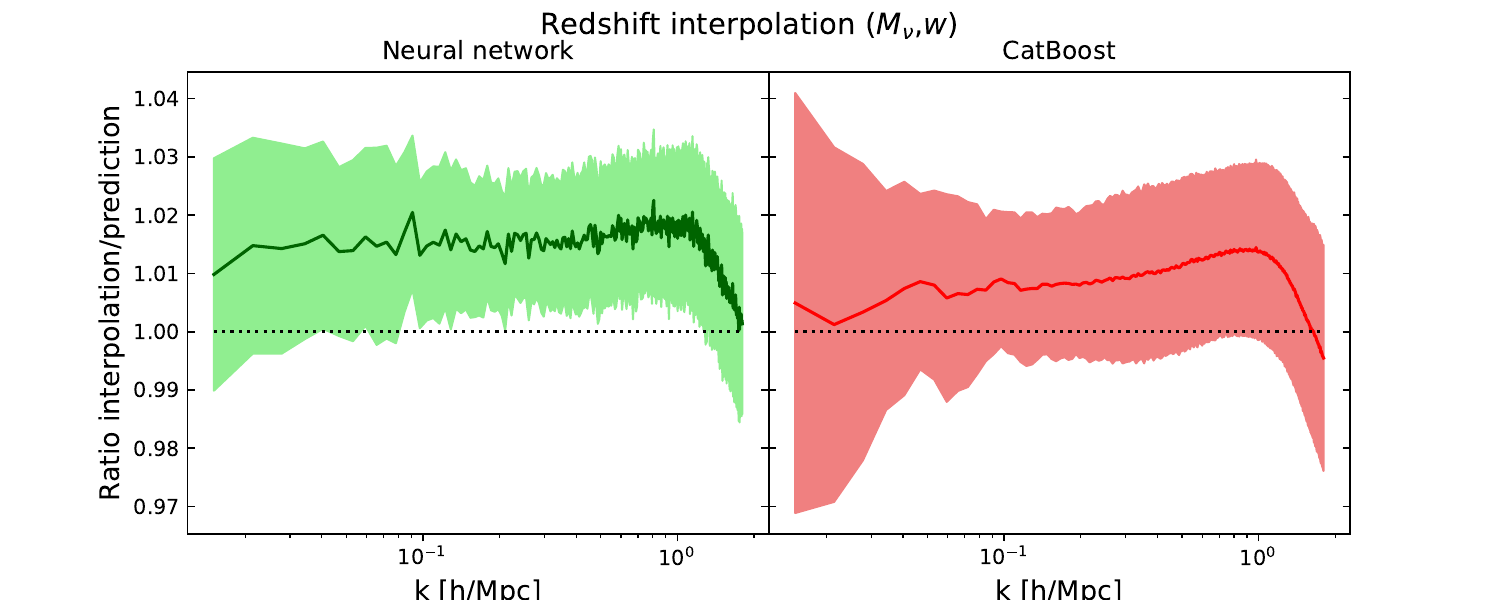} 
\caption{Comparison between the interpolated predictions at $z=1$ and the true predictions for the neural network and the CatBoost model. The solid line represents the mean of the ratios of the interpolated predictions and the model predictions at $z=1$, while the shaded areas represent $1\sigma$ error bars.}.
\label{fig:nu_interp}
\end{figure}

Due to the limited number of redshift snapshots available, we could not run inference models on the data. However, we know that for linear theory, the power spectrum can be expressed in terms of the growth rate $D(z)$ as $P_{\rm lin} (k,z)=D^2(z) P_{\rm lin} (k,0)$, where $D(z=0)=1$. This scaling function depends, of course, on the evolution history of the universe and hence on the cosmological parameters. However, one can see that $D^2(z)$ can be interpolated with a polynomial of degree 3 in redshift. Hence, for intermediate redshifts, we can use the outputs of the models for the 5 redshifts, and our emulator will output the interpolated values using least squares. To validate this approach, we have used redshifts 0, 0.5, 2 and 3 and we have predicted the power spectra at redshift 1 using cubic interpolation. The results are presented in Fig. \ref{fig:nu_interp}, where the ratio between the interpolation results and the model predictions are evaluated on the test set. The mean and the standard deviations are plotted for the neural network and the CatBoost model, showing a better than 5\% prediction accuracy through interpolation on the whole scale, reducing to less than $2.5\%$ for $k\ge0.1$. This accuracy is, however, limited, by the small number of redshifts where data is available. Hence, the emulator can be used for $0.1 \lesssim k \lesssim 2$ for $0 \le z \le 3$.

\section{Results for primordial non-Gaussianity}
\label{sec:results_fnl}
In this section we discuss the findings of the emulator involving $f_{\rm NL}^{\rm eq}$. In this case, there are five cosmological parameters that are varied, and hence the emulator will have 5-dimensional input data. The wavevector range is here $k \in [0.008900,5,569] \, h/{\rm Mpc}$ and we chose to restrict its range to $[0.015,2.0] \, h/{\rm Mpc}$. In these simulations, there are only four redshifts for the snapshots, and therefore the third-order fit for intermediate redshifts will be less precise and will not be performed in this work.

\begin{figure}[!h]
\centering
\includegraphics[width=0.75\linewidth]{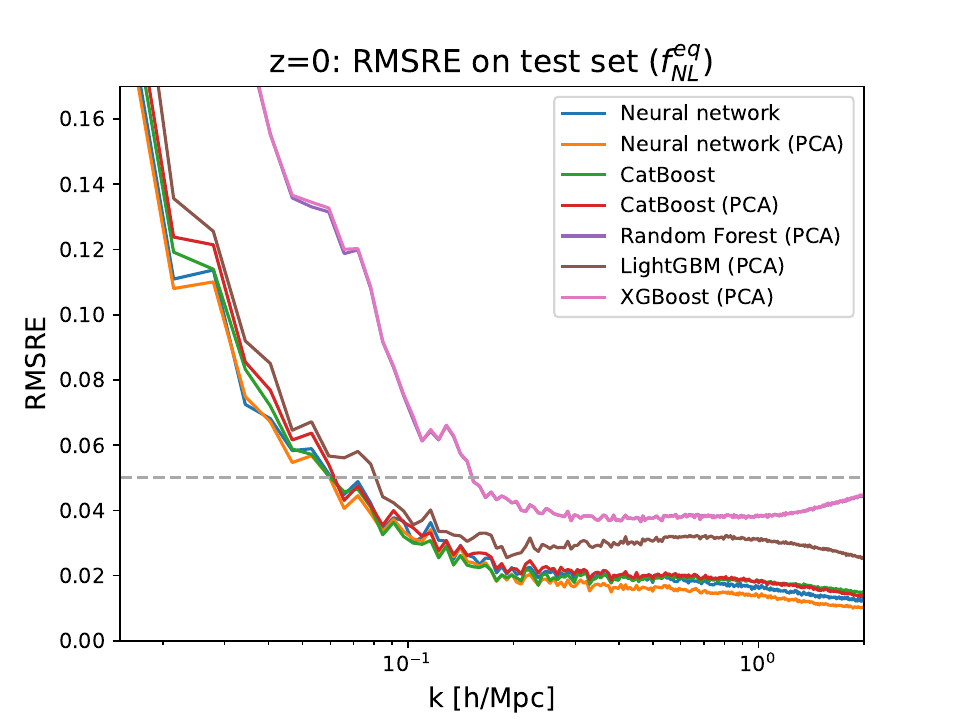} 
\caption{Comparison of RMSRE as a function of $k$ on the test set for the five-parameter emulator involving equilateral non-Gaussianity for the different learning models considered.}
\label{fig:fnlcomp}
\end{figure}

The number of simulations is 1000, and here we chose a random test set of size 100 to test the model. For convenience, we used a model architecture similar to the previous one, with two hidden layers for both the full power spectrum and the PCA versions, with the same number of neurons. In the case of the full power spectrum, there are 317 outputs, and for the PCA we again choose 11 principal components, which preserve 99.96\% of the variance. A comparison at $z=0$ of the RMSRE for the  models considered shows that the Random forests, LightGBM and XGBoost perform worse than the neutral network and CatBoost by a few percent on the test set. The neural network results are similar for the non-PCA and PCA versions, which is also the case for CatBoost, which performs marginally worse than the neural networks. However, we note that, in the case of this method, using PCA compression significantly reduces the training time and requires less tuning than a neural network, where the choice of architecture is essential. The running times of these models were similar to those of the first emulator.

The trends observed at $z=0$ are maintained at the other three redshifts as one can see from Table \ref{tab:rmsefnl}, where the overall RMSRE has been calculated. Indeed, the neural network (PCA compressed) works best at all redshifts, while CatBoost is slightly worse, with the two versions (PCA and non-PCA) almost indistinguishable.

\begin{table}[!h]
\begin{center}
\begin{tabular}{c|c|c|c|c}\hline
    Network & $z=0$ & $z=0.503$ & $z=0.733$ & $z=0.997$  \\ \hline\hline
    Neural network &144.3 & 99.4 & 80.4 & 68.1   \\
    \hline
    Neural network (PCA) &137.1& 91.4 & 78.1 & 63.7  \\
    \hline
    CatBoost & 145.8 & 99.4 & 84.3 & 69.2  \\
    \hline
    CatBoost (PCA) & 169.0 & 116.7 & 99.5 & 81.1   \\
    \hline
\end{tabular}
\caption{RMSEs for the 4 best models considered at the redshifts of the simulations (0, 0.503, 0.733, 0.997) for the emulator with five cosmological parameters: $\Omega_m$, $h$,  $n_s$, $\sigma_8$ and $f_{\rm NL}^{\rm eq}$.}
\label{tab:rmsefnl}
\end{center}
\end{table}

\begin{figure}[!h]
\centering
\includegraphics[width=0.75\linewidth]{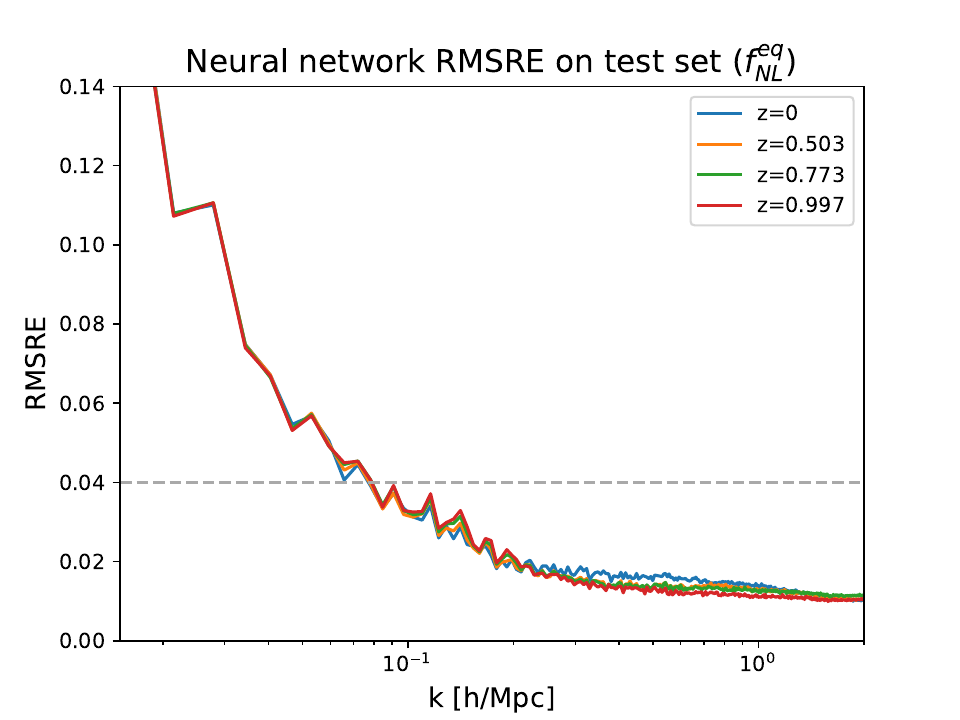} 
\caption{RMSRE for the neural network (with PCA) for redshifts 0, 0.503, 0.733 and 0.997 for the five-parameter emulator involving $f_{\rm NL}^{\rm eq}$.}
\label{fig:fnl_nn}
\end{figure}

To see the behaviour of the emulator at each scale, we again focussed on the best model -- the PCA compressed neural network. We plot the RMSRE in terms of scale for this model at the four redshifts considered (Fig. \ref{fig:fnl_nn}), and we see that this metric is lower than 4\% for $k>0.1\, h/{\rm Mpc}$ and this drops to around 1\% for $k \sim 2\, h/{\rm Mpc}$. For $k<0.1\, h/{\rm Mpc}$, the results are not so accurate, due to the cosmic variance that has not been mitigated by running a large number of simulations. Hence, this emulator has an accuracy of less than 4\% for $0.1 \lesssim k \lesssim 2$ for $0 \le z \le 3$.

\section{Performance of the emulator and comparison with other emulators}
\label{sec:comparisons}

To evaluate the performance of our dark energy emulator, and to show the practical benefit of the emulators, we performed a MCMC \cite{Hastings1970} analysis of the 7-parameter emulator and we compared its performance to \texttt{HALOFIT}. In the analysis we have employed the emulator based on CatBoost. For simplicity we chose a fiducial cosmology, whose power spectrum has been computed with \texttt{HALOFIT}, and we have employed the Python \texttt{emcee} package \cite{Foreman-Mackey:2012any} to perform the MCMC analysis. We have used a diagonal covariance matrix in the likelihood function, proportional to the power spectrum. The priors were chosen to correspond to the parameter space of the training of the emulator. We ran the MCMC analysis for both the emulator and \texttt{HALOFIT}, focussing on the range $0.1 \le k \le 1 h/\rm{Mpc}$ at $z=0$. The results, including the fiducial values of the parameters, posterior mean and $1\sigma$ error bars are presented in Table \ref{tab:emul_vs_hf}. We also show the execution time of the code, which highlights the significant time saving (of a factor of $\mathcal{O}(2500)$) with respect to \texttt{HALOFIT} that becomes essential in situations where the non-linear matter power spectrum must be determined a large number of times for different parameter values. 

\begin{table}[!h]
\begin{center}
\begin{tabular}{c|c|c|c}\hline
Parameter & Fiducial & Emulator & \texttt{HALOFIT} \\ \hline\hline
$\Omega_m$ & 0.3 & $0.299\pm0.024$ & $0.301 \pm 0.010$  \\ \hline
$\Omega_b$ & 0.048 & $0.05016\pm0.00920$  & $0.0478 \pm 0.00887$ \\ \hline
\textit{h} & 0.67 & $0.6787\pm0.0650$ & $0.6835 \pm 0.0630$ \\ \hline
$\sigma_8$ & 0.82 & $0.8131 \pm0.0108$ & $0.8200 \pm 0.0115$ \\ \hline
$n_s$ & 0.965 & $0.95282\pm0.03370$ & $0.9609 \pm 0.0347$ \\ \hline
$M_{\nu}$ & 0.4 & $0.472\pm0.272$ & $0.427 \pm 0.223$ \\ \hline
\textit{w} & $-0.85$ & $-0.9377\pm0.1452$ & $-0.8754 \pm 0.1221$ \\ \hline \hline
Wall time (s) & -- & 19 & 51431 \\ \hline \hline
\hline
\end{tabular}
\caption{Comparison between the fits obtained through the MCMC procedure for the emulator and \texttt{HALOFIT}, showing the mean and standard deviations for each of the parameters, compared to the fiducial values used (left column). The wall time in seconds for the emulator and \texttt{HALOFIT} are showed in the last row, illustrating the significant improvement in the running time of the emulator.}
\label{tab:emul_vs_hf}
\end{center}
\end{table}

\begin{figure}[!h]
\centering
\includegraphics[width=
\linewidth]{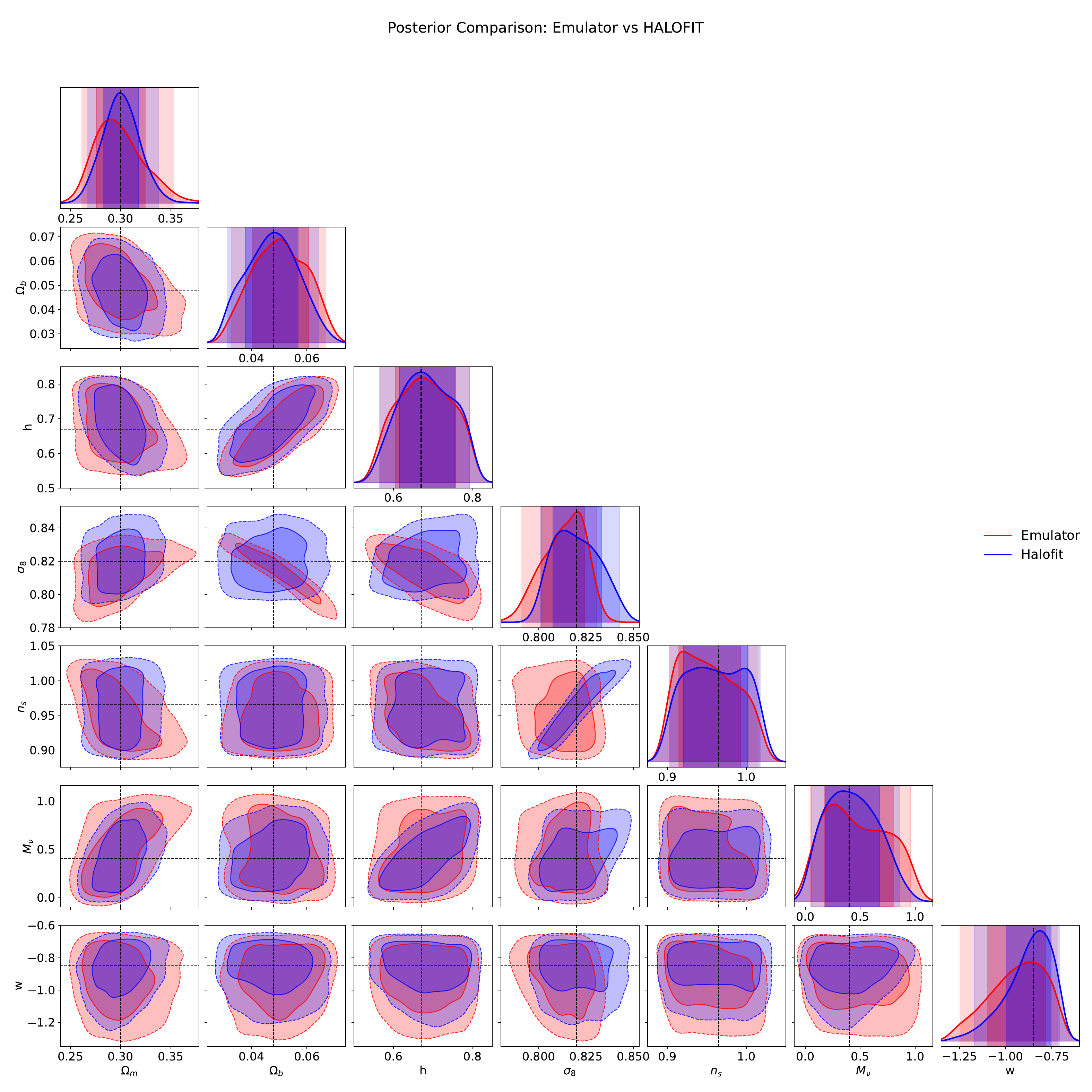} 
\caption{Posterior distributions of cosmological parameters comparing the Emulator  (red) and the \texttt{HALOFIT} model (blue). Diagonal panels show marginalized one-dimensional posteriors for each parameter, with shaded regions denoting the 68\% and 95\% credible intervals. Off-diagonal panels display the two-dimensional joint posteriors, where filled contours correspond to the $1\sigma$ and $2\sigma$ credible regions, derived from Gaussian kernel density estimates. Red contours represent the Emulator results, blue contours represent \texttt{HALOFIT}, and dashed black lines mark the fiducial parameter values.}
\label{fig:comp_emu_hf}
\end{figure}

In Fig. \ref{fig:comp_emu_hf} we show the marginalised histograms for each of the parameters, the fiducial value (dashed) and 1$\sigma$ and 2$\sigma$ error bars for the emulator (red) and \texttt{HALOFIT} (blue). Constraints for parameter pairs are also represented, with 1$\sigma$ and 2$\sigma$ contours and the fiducial values (intersection of the dashed lines). This analysis confirms in a practical setting that the emulator is accurate and that it significantly improves the speed with respect to \texttt{HALOFIT}. We note, however, the large error bars for the $w$ and $M_\nu$ parameters, corresponding to degeneracies between these parameters that cannot be broken using the matter power spectrum.

We would like to mention some emulators built by other groups, in particular \texttt{CosmoPower} \cite{SpurioMancini:2021ppk}, \texttt{CosmoPower-JAX} \cite{Piras:2023aub}, \texttt{EuclidEmulator2} \cite{Euclid:2020rfv}, \texttt{BACCO} \cite{Angulo:2020vky}, \texttt{syren-new} \cite{Sui:2024wob} and \texttt{FREmu} \cite{Bai:2024cgt}. 
\texttt{CosmoPower} and \texttt{CosmoPower-JAX} investigate $\Lambda$CDM cosmologies in the range $k \in [10^{-5},10]\;\rm{Mpc}^{-1}$ and $0\le z \le 5$ and have more than 99\% accuracy. The \texttt{EuclidEmulator2}, \texttt{BACCO} and \texttt{syren-new} emulators can predict in addition the dark energy equation of state parameters (parametrised via $w(a)=w_0+w_a(1-a)$) and the total neutrino mass and have accuracies of 99\%, 97\% and 98.7\%, on scales of [0.01,10] \textit{h}/Mpc, [0.01,5] \textit{h}/Mpc and $[9\times 10^{-3},9]\; h/{\rm Mpc}$ and redshifts less than 3, 1.5 and 3 respectively.

Finally \texttt{FREmu} has been tuned on $f(R)$ modified gravity theories. We note that the authors employ simulations from the \textsc{Quijote} suite, related to the ones used in this work, and they obtain a similar accuracy in their results, for $0<z<3$ and $k \in [0.0089,0.5]\; h/{\rm Mpc}$.

We note that our emulator is the only one that investigates equilateral primordial non-Gaussianity.

\section{Discussion, conclusions and future directions}
\label{sec:conc}

We have shown how accurate $N$-body \textsc{Quijote} simulations can be used to build power spectrum emulators for different cosmological parameters. In particular, we built two power spectrum emulators: one varying seven cosmological parameters, including the total sum of neutrino masses and the parameter describing the equation of state of dark energy, and a second one, varying five parameters including the amplitude of primordial equilateral non-Gaussianity.

In this work we have made the choice to look at the logarithm of the power spectrum from the simulation rather than consider the ratio of the model to the linear or some nonlinear (like \texttt{HALOFIT}) version of the matter power spectrum which has been used in the literature. This procedure has the advantage that it does not require external data, or running a Boltzmann code, to generate the linear or non-linear power spectrum.

For each of the emulators we have investigated several machine learning models, including neural networks and tree-based methods. We have looked into methods to compress the power spectrum outputs and we showed that 11 principal components from PCA can encode more than 99.5\% of the variance of the data. 

In the case of the neural networks, the best performing model determined has an architecture with two 2048-neuron hidden layers with ReLU activation functions, followed by Dropout layers. The dropout rates are of 15\% and 60\%
for the full and PCA versions respectively. 
For both emulators, the PCA neural network achieved better than 5\% RMSRE for all the scales considered. In what concerns tree-based methods, we have investigated Random Forests,  CatBoost,  XGBoost  and LightGBM, showing that only CatBoost can provide competitive results to the neural networks, with RMSRE less than 1\% worse than the them, with an overall RMSRE error at the 3-4\% level for $0.1 \lesssim k \lesssim 2$ and $0 \le z \le 3$. For intermediate redshifts, one can numerically interpolate using a third-order polynomial in redshifts. 

By performing a MCMC analysis, we confirmed the above results and also a similar constraining power to \texttt{HALOFIT}. It also highlighted the significant time saving with respect to the latter, the emulator being roughly 2500 faster.

The results that we have obtained here can be improved by significantly increasing the number of simulations used for training the networks. 
As emulators are powerful tools that can be used to replace simulations in situations where a large number of runs is required, we aim to investigate this area in more detail in future work, and in particular to look at different cosmological quantities, such as the bispectrum.

\section*{Acknowledgments}
The author has been supported by a United Kingdom
Research and Innovation (UKRI) Future Leaders Fellowship [Grant
No. MR/V021974/2] and by the Romanian Ministry of Research, Innovation and Digitalization under the Romanian National Core Program LAPLAS VII-contract no. 30N/2023.

\section*{Data availability}
The learning models are available as Jupyter notebooks on GitHub at \url{https://github.com/andreilazanu/power_spectrum_emulators}. The data from the \textsc{Quijote} simulations is available on \url{https://quijote-simulations.readthedocs.io/en/latest/}, where detailed instructions are given on how to download and use the information available.

\bibliographystyle{elsarticle-num-names} 
\bibliography{Bibliography}

@article{10.2307/1268522,
 ISSN = {00401706},
 URL = {http://www.jstor.org/stable/1268522},
 author = {M. D. McKay and R. J. Beckman and W. J. Conover},
 journal = {Technometrics},
 number = {2},
 pages = {239--245},
 publisher = {[Taylor & Francis, Ltd., American Statistical Association, American Society for Quality]},
 title = {A Comparison of Three Methods for Selecting Values of Input Variables in the Analysis of Output from a Computer Code},
 volume = {21},
 year = {1979}
}

@article{Collister:2003cz,
    author = "Collister, Adrian A. and Lahav, Ofer",
    title = "{ANNz: Estimating photometric redshifts using artificial neural networks}",
    eprint = "astro-ph/0311058",
    archivePrefix = "arXiv",
    doi = "10.1086/383254",
    journal = "Publ. Astron. Soc. Pac.",
    volume = "116",
    pages = "345--351",
    year = "2004"
}

@article{Eriksen:2020diu,
    author = "Eriksen, M. and others",
    title = "{The PAU Survey: Photometric redshifts using transfer learning from simulations}",
    eprint = "2004.07979",
    archivePrefix = "arXiv",
    primaryClass = "astro-ph.GA",
    doi = "10.1093/mnras/staa2265",
    journal = "Mon. Not. Roy. Astron. Soc.",
    volume = "497",
    number = "4",
    pages = "4565--4579",
    year = "2020"
}

@article{Lin:2021dim,
    author = "Lin, Z. and Huang, N. and Avestruz, C. and Wu, W. L. K. and Trivedi, S. and Caldeira, J. and Nord, B.",
    title = "{DeepSZ: Identification of Sunyaev-Zel'dovich Galaxy Clusters using Deep Learning}",
    eprint = "2102.13123",
    archivePrefix = "arXiv",
    primaryClass = "astro-ph.CO",
    reportNumber = "FERMILAB-PUB-21-077-SCD",
    month = "2",
    year = "2021"
}

@article{Rodriguez:2018mjb,
    author = "Rodriguez, Andres C. and Kacprzak, Tomasz and Lucchi, Aurelien and Amara, Adam and Sgier, Raphael and Fluri, Janis and Hofmann, Thomas and R\'efr\'egier, Alexandre",
    title = "{Fast cosmic web simulations with generative adversarial networks}",
    eprint = "1801.09070",
    archivePrefix = "arXiv",
    primaryClass = "astro-ph.CO",
    doi = "10.1186/s40668-018-0026-4",
    journal = "Comput. Astrophys. Cosmol.",
    volume = "5",
    pages = "4",
    year = "2018"
}

@article{Xu:2021pdp,
    author = "Xu, Xiaoju and Kumar, Saurabh and Zehavi, Idit and Contreras, Sergio",
    title = "{Predicting halo occupation and galaxy assembly bias with machine learning}",
    eprint = "2107.01223",
    archivePrefix = "arXiv",
    primaryClass = "astro-ph.CO",
    month = "7",
    year = "2021"
}

@article{Lucie-Smith:2020ris,
    author = "Lucie-Smith, Luisa and Peiris, Hiranya V. and Pontzen, Andrew and Nord, Brian and Thiyagalingam, Jeyan",
    title = "{Deep learning insights into cosmological structure formation}",
    eprint = "2011.10577",
    archivePrefix = "arXiv",
    primaryClass = "astro-ph.CO",
    reportNumber = "FERMILAB-PUB-20-643-SCD",
    month = "11",
    year = "2020"
}

@article{Huang:2020wvt,
    author = "Huang, Lawrence and Croft, Rupert A. C. and Arora, Hitesh",
    title = "{Deep Forest: Neural Network reconstruction of the Lyman-alpha forest}",
    eprint = "2009.10673",
    archivePrefix = "arXiv",
    primaryClass = "astro-ph.CO",
    month = "9",
    year = "2020"
}

@article{Shimabukuro:2017jdh,
    author = "Shimabukuro, Hayato and Semelin, Benoit",
    title = "{Analysing the 21 cm signal from the epoch of reionization with artificial neural networks}",
    eprint = "1701.07026",
    archivePrefix = "arXiv",
    primaryClass = "astro-ph.CO",
    doi = "10.1093/mnras/stx734",
    journal = "Mon. Not. Roy. Astron. Soc.",
    volume = "468",
    number = "4",
    pages = "3869--3877",
    year = "2017"
}

@article{ZorrillaMatilla:2020doz,
    author = "Zorrilla Matilla, Jos\'e Manuel and Sharma, Manasi and Hsu, Daniel and Haiman, Zolt\'an",
    title = "{Interpreting deep learning models for weak lensing}",
    eprint = "2007.06529",
    archivePrefix = "arXiv",
    primaryClass = "astro-ph.CO",
    doi = "10.1103/PhysRevD.102.123506",
    journal = "Phys. Rev. D",
    volume = "102",
    number = "12",
    pages = "123506",
    year = "2020"
}

@article{Ribli:2018kwb,
    author = "Ribli, Dezs\H{o} and Pataki, B\'alint \'Armin and Csabai, Istv\'an",
    title = "{An improved cosmological parameter inference scheme motivated by deep learning}",
    eprint = "1806.05995",
    archivePrefix = "arXiv",
    primaryClass = "astro-ph.CO",
    doi = "10.1038/s41550-018-0596-8",
    journal = "Nature Astron.",
    volume = "3",
    number = "1",
    pages = "93--98",
    year = "2019"
}

@article{Park:2020eat,
    author = "Park, Ji Won and Wagner-Carena, Sebastian and Birrer, Simon and Marshall, Philip J. and Lin, Joshua Yao-Yu and Roodman, Aaron",
    collaboration = "LSST Dark Energy Science",
    title = "{Large-Scale Gravitational Lens Modeling with Bayesian Neural Networks for Accurate and Precise Inference of the Hubble Constant}",
    eprint = "2012.00042",
    archivePrefix = "arXiv",
    primaryClass = "astro-ph.IM",
    doi = "10.3847/1538-4357/abdfc4",
    journal = "Astrophys. J.",
    volume = "910",
    number = "1",
    pages = "39",
    year = "2021"
}

@article{Jacobs:2017xhn,
    author = "Jacobs, Colin and Glazebrook, Karl and Collett, Thomas and More, Anupreeta and McCarthy, Christopher",
    title = "{Finding strong lenses in CFHTLS using convolutional neural networks}",
    eprint = "1704.02744",
    archivePrefix = "arXiv",
    primaryClass = "astro-ph.IM",
    doi = "10.1093/mnras/stx1492",
    journal = "Mon. Not. Roy. Astron. Soc.",
    volume = "471",
    number = "1",
    pages = "167--181",
    year = "2017"
}

@article{Chanda:2021qbf,
    author = "Chanda, Pallav and Saha, Rajib",
    title = "{An Unbiased Estimator of the Full-sky CMB Angular Power Spectrum using Neural Networks}",
    eprint = "2102.04327",
    archivePrefix = "arXiv",
    primaryClass = "astro-ph.CO",
    month = "2",
    year = "2021"
}

@article{Caldeira:2018ojb,
    author = "Caldeira, Jo\~ao and Wu, W. L. Kimmy and Nord, Brian and Avestruz, Camille and Trivedi, Shubhendu and Story, Kyle T.",
    title = "{DeepCMB: Lensing Reconstruction of the Cosmic Microwave Background with Deep Neural Networks}",
    eprint = "1810.01483",
    archivePrefix = "arXiv",
    primaryClass = "astro-ph.CO",
    reportNumber = "FERMILAB-PUB-18-515-A-CD",
    doi = "10.1016/j.ascom.2019.100307",
    journal = "Astron. Comput.",
    volume = "28",
    pages = "100307",
    year = "2019"
}

@article{Alsing:2019xrx,
    author = "Alsing, Justin and Charnock, Tom and Feeney, Stephen and Wandelt, Benjamin",
    title = "{Fast likelihood-free cosmology with neural density estimators and active learning}",
    eprint = "1903.00007",
    archivePrefix = "arXiv",
    primaryClass = "astro-ph.CO",
    doi = "10.1093/mnras/stz1960",
    journal = "Mon. Not. Roy. Astron. Soc.",
    volume = "488",
    number = "3",
    pages = "4440--4458",
    year = "2019"
}

@article{Kostic:2021tyw,
    author = "Kosti\'c, Andrija and Jasche, Jens and Ramanah, Doogesh Kodi and Lavaux, Guilhem",
    title = "{Machine-driven searches for cosmological physics}",
    eprint = "2107.00657",
    archivePrefix = "arXiv",
    primaryClass = "astro-ph.CO",
    month = "7",
    year = "2021"
}

@ARTICLE{2003ApJ...583....1B,
       author = {{Bennett}, C.~L. and {Bay}, M. and {Halpern}, M. and {Hinshaw}, G. and {Jackson}, C. and {Jarosik}, N. and {Kogut}, A. and {Limon}, M. and {Meyer}, S.~S. and {Page}, L. and {Spergel}, D.~N. and {Tucker}, G.~S. and {Wilkinson}, D.~T. and {Wollack}, E. and {Wright}, E.~L.},
        title = "{The Microwave Anisotropy Probe Mission}",
      journal = {\apj},
         year = 2003,
        month = jan,
       volume = {583},
       number = {1},
        pages = {1-23},
          doi = {10.1086/345346},
archivePrefix = {arXiv},
       eprint = {astro-ph/0301158},
 primaryClass = {astro-ph},
       adsurl = {https://ui.adsabs.harvard.edu/abs/2003ApJ...583....1B}
}

@article{Crocce:2006ve,
    author = "Crocce, M. and Pueblas, S. and Scoccimarro, R.",
    title = "{Transients from Initial Conditions in Cosmological Simulations}",
    eprint = "astro-ph/0606505",
    archivePrefix = "arXiv",
    doi = "10.1111/j.1365-2966.2006.11040.x",
    journal = "Mon. Not. Roy. Astron. Soc.",
    volume = "373",
    pages = "369--381",
    year = "2006"
}

@article{Scoccimarro:1997gr,
    author = "Scoccimarro, Roman",
    title = "{Transients from initial conditions: a perturbative analysis}",
    eprint = "astro-ph/9711187",
    archivePrefix = "arXiv",
    reportNumber = "CITA-97-42",
    doi = "10.1046/j.1365-8711.1998.01845.x",
    journal = "Mon. Not. Roy. Astron. Soc.",
    volume = "299",
    pages = "1097",
    year = "1998"
}

@article{Lewis:1999bs,
    author = "Lewis, Antony and Challinor, Anthony and Lasenby, Anthony",
    title = "{Efficient computation of CMB anisotropies in closed FRW models}",
    eprint = "astro-ph/9911177",
    archivePrefix = "arXiv",
    doi = "10.1086/309179",
    journal = "Astrophys. J.",
    volume = "538",
    pages = "473--476",
    year = "2000"
}

@article{Springel:2005mi,
    author = "Springel, Volker",
    title = "{The Cosmological simulation code GADGET-2}",
    eprint = "astro-ph/0505010",
    archivePrefix = "arXiv",
    doi = "10.1111/j.1365-2966.2005.09655.x",
    journal = "Mon. Not. Roy. Astron. Soc.",
    volume = "364",
    pages = "1105--1134",
    year = "2005"
}

@article{Jarvis:2015tqa,
    author = "Jarvis, Matt J. and Bacon, David and Blake, Chris and Brown, Michael L. and Lindsay, Sam N. and Raccanelli, Alvise and Santos, Mario and Schwarz, Dominik",
    title = "{Cosmology with SKA Radio Continuum Surveys}",
    eprint = "1501.03825",
    archivePrefix = "arXiv",
    primaryClass = "astro-ph.CO",
    month = "1",
    year = "2015"
}

@article{Laureijs:2011gra,
    author = "Laureijs, R. and others",
    collaboration = "EUCLID",
    title = "{Euclid Definition Study Report}",
    eprint = "1110.3193",
    archivePrefix = "arXiv",
    primaryClass = "astro-ph.CO",
    reportNumber = "ESA-SRE(2011)12",
    month = "10",
    year = "2011"
}

@article{Abell:2009aa,
    author = "Abell, Paul A. and others",
    collaboration = "LSST Science, LSST Project",
    title = "{LSST Science Book, Version 2.0}",
    eprint = "0912.0201",
    archivePrefix = "arXiv",
    primaryClass = "astro-ph.IM",
    reportNumber = "FERMILAB-TM-2495-A, SLAC-R-1031",
    month = "12",
    year = "2009"
}

@article{Akrami:2018vks,
    author = "Aghanim, N. and others",
    collaboration = "Planck",
    title = "{Planck 2018 results. I. Overview and the cosmological legacy of Planck}",
    eprint = "1807.06205",
    archivePrefix = "arXiv",
    primaryClass = "astro-ph.CO",
    doi = "10.1051/0004-6361/201833880",
    journal = "Astron. Astrophys.",
    volume = "641",
    pages = "A1",
    year = "2020"
}

@ARTICLE{Quijote_sims,
         author = {{Villaescusa-Navarro}, Francisco and {Hahn}, ChangHoon and {Massara}, Elena and {Banerjee}, Arka and {Delgado}, Ana Maria and {Ramanah}, Doogesh Kodi and {Charnock}, Tom and {Giusarma}, Elena and {Li}, Yin and {Allys}, Erwan and {Brochard}, Antoine and {Uhlemann}, Cora and {Chiang}, Chi-Ting and {He}, Siyu and {Pisani}, Alice and {Obuljen}, Andrej and {Feng}, Yu and {Castorina}, Emanuele and {Contardo}, Gabriella and {Kreisch}, Christina D. and {Nicola}, Andrina and {Alsing}, Justin and {Scoccimarro}, Roman and {Verde}, Licia and {Viel}, Matteo and {Ho}, Shirley and {Mallat}, Stephane and {Wandelt}, Benjamin and {Spergel}, David N.},
         title = "{The Quijote Simulations}",
         journal = {\apjs},
         year = 2020,
         month = sep,
         volume = {250},
         number = {1},
         eid = {2},
         pages = {2},
         doi = {10.3847/1538-4365/ab9d82},
         archivePrefix = {arXiv},
         eprint = {1909.05273},
         primaryClass = {astro-ph.CO},
         adsurl = {https://ui.adsabs.harvard.edu/abs/2020ApJS..250....2V}
}

@article{RevModPhys.61.1,
  title = {The cosmological constant problem},
  author = {Weinberg, Steven},
  journal = {Rev. Mod. Phys.},
  volume = {61},
  issue = {1},
  pages = {1--23},
  numpages = {0},
  year = {1989},
  month = {Jan},
  publisher = {American Physical Society},
  doi = {10.1103/RevModPhys.61.1},
  url = {https://link.aps.org/doi/10.1103/RevModPhys.61.1}
}

@article{Zennaro:2016nqo,
    author = "Zennaro, Matteo and Bel, Julien and Villaescusa-Navarro, Francisco and Carbone, Carmelita and Sefusatti, Emiliano and Guzzo, Luigi",
    title = "{Initial Conditions for Accurate N-Body Simulations of Massive Neutrino Cosmologies}",
    eprint = "1605.05283",
    archivePrefix = "arXiv",
    primaryClass = "astro-ph.CO",
    doi = "10.1093/mnras/stw3340",
    journal = "Mon. Not. Roy. Astron. Soc.",
    volume = "466",
    number = "3",
    pages = "3244--3258",
    year = "2017"
}

@article{Pearson01111901,
author = {Karl Pearson},
title = {LIII. On lines and planes of closest fit to systems of points in space },
journal = {The London, Edinburgh, and Dublin Philosophical Magazine and Journal of Science},
volume = {2},
number = {11},
pages = {559--572},
year = {1901},
publisher = {Taylor \& Francis},
doi = {10.1080/14786440109462720}
}

@misc{CatBoost,
      title={CatBoost: gradient boosting with categorical features support}, 
      author={Anna Veronika Dorogush and Vasily Ershov and Andrey Gulin},
      year={2018},
      eprint={1810.11363},
      archivePrefix={arXiv},
      primaryClass={cs.LG},
      url={https://arxiv.org/abs/1810.11363}, 
}

@article{Habib:2007ca,
    author = "Habib, Salman and Heitmann, Katrin and Higdon, David and Nakhleh, Charles and Williams, Brian",
    title = "{Cosmic Calibration: Constraints from the Matter Power Spectrum and the Cosmic Microwave Background}",
    eprint = "astro-ph/0702348",
    archivePrefix = "arXiv",
    reportNumber = "LA-UR-07-0056",
    doi = "10.1103/PhysRevD.76.083503",
    journal = "Phys. Rev. D",
    volume = "76",
    pages = "083503",
    year = "2007"
}

@article{Heitmann:2013bra,
    author = "Heitmann, Katrin and Lawrence, Earl and Kwan, Juliana and Habib, Salman and Higdon, David",
    title = "{The Coyote Universe Extended: Precision Emulation of the Matter Power Spectrum}",
    eprint = "1304.7849",
    archivePrefix = "arXiv",
    primaryClass = "astro-ph.CO",
    reportNumber = "ANL-HEP-PR-13-10",
    doi = "10.1088/0004-637X/780/1/111",
    journal = "Astrophys. J.",
    volume = "780",
    pages = "111",
    year = "2014"
}

@article{Heitmann:2008eq,
    author = "Heitmann, Katrin and White, Martin and Wagner, Christian and Habib, Salman and Higdon, David",
    title = "{The Coyote Universe I: Precision Determination of the Nonlinear Matter Power Spectrum}",
    eprint = "0812.1052",
    archivePrefix = "arXiv",
    primaryClass = "astro-ph",
    reportNumber = "LA-UR-08-05630",
    doi = "10.1088/0004-637X/715/1/104",
    journal = "Astrophys. J.",
    volume = "715",
    pages = "104--121",
    year = "2010"
}

@article{Heitmann:2009cu,
    author = "Heitmann, Katrin and Higdon, David and White, Martin and Habib, Salman and Williams, Brian J. and Wagner, Christian",
    title = "{The Coyote Universe II: Cosmological Models and Precision Emulation of the Nonlinear Matter Power Spectrum}",
    eprint = "0902.0429",
    archivePrefix = "arXiv",
    primaryClass = "astro-ph.CO",
    reportNumber = "LA-UR-08-07921",
    doi = "10.1088/0004-637X/705/1/156",
    journal = "Astrophys. J.",
    volume = "705",
    pages = "156--174",
    year = "2009"
}

@article{Lawrence:2009uk,
    author = "Lawrence, Earl and Heitmann, Katrin and White, Martin and Higdon, David and Wagner, Christian and Habib, Salman and Williams, Brian",
    title = "{The Coyote Universe III: Simulation Suite and Precision Emulator for the Nonlinear Matter Power Spectrum}",
    eprint = "0912.4490",
    archivePrefix = "arXiv",
    primaryClass = "astro-ph.CO",
    reportNumber = "LA-UR-09-06131",
    doi = "10.1088/0004-637X/713/2/1322",
    journal = "Astrophys. J.",
    volume = "713",
    pages = "1322--1331",
    year = "2010"
}

@article{Heitmann:2006hr,
    author = "Heitmann, Katrin and Higdon, David and Nakhleh, Charles and Habib, Salman",
    title = "{Cosmic Calibration}",
    eprint = "astro-ph/0606154",
    archivePrefix = "arXiv",
    reportNumber = "LA-UR-06-2320",
    doi = "10.1086/506448",
    journal = "Astrophys. J. Lett.",
    volume = "646",
    pages = "L1--L4",
    year = "2006"
}

@article{Agarwal:2012ew,
    author = "Agarwal, Shankar and Abdalla, Filipe B. and Feldman, Hume A. and Lahav, Ofer and Thomas, Shaun A.",
    title = "{PkANN - I. Non-linear matter power spectrum interpolation through artificial neural networks}",
    eprint = "1203.1695",
    archivePrefix = "arXiv",
    primaryClass = "astro-ph.CO",
    doi = "10.1111/j.1365-2966.2012.21326.x",
    journal = "Mon. Not. Roy. Astron. Soc.",
    volume = "424",
    pages = "1409--1418",
    year = "2012"
}

@article{Smith:2002dz,
    author = "Smith, R. E. and Peacock, J. A. and Jenkins, A. and White, S. D. M. and Frenk, C. S. and Pearce, F. R. and Thomas, P. A. and Efstathiou, G. and Couchmann, H. M. P.",
    collaboration = "VIRGO Consortium",
    title = "{Stable clustering, the halo model and nonlinear cosmological power spectra}",
    eprint = "astro-ph/0207664",
    archivePrefix = "arXiv",
    doi = "10.1046/j.1365-8711.2003.06503.x",
    journal = "Mon. Not. Roy. Astron. Soc.",
    volume = "341",
    pages = "1311",
    year = "2003"
}

@article{Takahashi:2012em,
    author = "Takahashi, Ryuichi and Sato, Masanori and Nishimichi, Takahiro and Taruya, Atsushi and Oguri, Masamune",
    title = "{Revising the Halofit Model for the Nonlinear Matter Power Spectrum}",
    eprint = "1208.2701",
    archivePrefix = "arXiv",
    primaryClass = "astro-ph.CO",
    doi = "10.1088/0004-637X/761/2/152",
    journal = "Astrophys. J.",
    volume = "761",
    pages = "152",
    year = "2012"
}

@article{Agarwal:2013aea,
    author = "Agarwal, Shankar and Abdalla, Filipe B. and Feldman, Hume A. and Lahav, Ofer and Thomas, Shaun A.",
    title = "{pkann \textendash{} II. A non-linear matter power spectrum interpolator developed using artificial neural networks}",
    eprint = "1312.2101",
    archivePrefix = "arXiv",
    primaryClass = "astro-ph.CO",
    doi = "10.1093/mnras/stu090",
    journal = "Mon. Not. Roy. Astron. Soc.",
    volume = "439",
    number = "2",
    pages = "2102--2121",
    year = "2014"
}

@article{Mohammed:2014lja,
    author = "Mohammed, Irshad and Seljak, Uros",
    title = "{Analytic model for the matter power spectrum, its covariance matrix, and baryonic effects}",
    eprint = "1407.0060",
    archivePrefix = "arXiv",
    primaryClass = "astro-ph.CO",
    doi = "10.1093/mnras/stu1972",
    journal = "Mon. Not. Roy. Astron. Soc.",
    volume = "445",
    number = "4",
    pages = "3382--3400",
    year = "2014"
}

@article{Winther:2019mus,
    author = "Winther, Hans and Casas, Santiago and Baldi, Marco and Koyama, Kazuya and Li, Baojiu and Lombriser, Lucas and Zhao, Gong-Bo",
    title = "{Emulators for the nonlinear matter power spectrum beyond $\Lambda$CDM}",
    eprint = "1903.08798",
    archivePrefix = "arXiv",
    primaryClass = "astro-ph.CO",
    doi = "10.1103/PhysRevD.100.123540",
    journal = "Phys. Rev. D",
    volume = "100",
    number = "12",
    pages = "123540",
    year = "2019"
}

@article{Ramanah:2020vyl,
    author = "Ramanah, Doogesh Kodi and Charnock, Tom and Villaescusa-Navarro, Francisco and Wandelt, Benjamin D.",
    title = "{Super-resolution emulator of cosmological simulations using deep physical models}",
    eprint = "2001.05519",
    archivePrefix = "arXiv",
    primaryClass = "astro-ph.CO",
    doi = "10.1093/mnras/staa1428",
    journal = "Mon. Not. Roy. Astron. Soc.",
    volume = "495",
    pages = "4227",
    year = "2020"
}

@article{Angulo:2020vky,
    author = {Angulo, Raul E. and Zennaro, Matteo and Contreras, Sergio and Aric\`o, Giovanni and Pellejero-Iba\~nez, Marcos and St\"ucker, Jens},
    title = "{The BACCO simulation project: exploiting the full power of large-scale structure for cosmology}",
    eprint = "2004.06245",
    archivePrefix = "arXiv",
    primaryClass = "astro-ph.CO",
    doi = "10.1093/mnras/stab2018",
    journal = "Mon. Not. Roy. Astron. Soc.",
    volume = "507",
    number = "4",
    pages = "5869--5881",
    year = "2021"
}

@article{Mead:2020vgs,
    author = {Mead, Alexander and Brieden, Samuel and Tr\"oster, Tilman and Heymans, Catherine},
    title = "{hmcode-2020: improved modelling of non-linear cosmological power spectra with baryonic feedback}",
    eprint = "2009.01858",
    archivePrefix = "arXiv",
    primaryClass = "astro-ph.CO",
    doi = "10.1093/mnras/stab082",
    journal = "Mon. Not. Roy. Astron. Soc.",
    volume = "502",
    number = "1",
    pages = "1401--1422",
    year = "2021"
}

@article{Ramachandra:2020lue,
    author = "Ramachandra, Nesar and Valogiannis, Georgios and Ishak, Mustapha and Heitmann, Katrin",
    collaboration = "LSST Dark Energy Science",
    title = "{Matter Power Spectrum Emulator for f(R) Modified Gravity Cosmologies}",
    eprint = "2010.00596",
    archivePrefix = "arXiv",
    primaryClass = "astro-ph.CO",
    doi = "10.1103/PhysRevD.103.123525",
    journal = "Phys. Rev. D",
    volume = "103",
    number = "12",
    pages = "123525",
    year = "2021"
}

@article{Euclid:2020rfv,
    author = "Knabenhans, M. and others",
    collaboration = "Euclid",
    title = "{Euclid preparation: IX. EuclidEmulator2 \textendash{} power spectrum emulation with massive neutrinos and self-consistent dark energy perturbations}",
    eprint = "2010.11288",
    archivePrefix = "arXiv",
    primaryClass = "astro-ph.CO",
    doi = "10.1093/mnras/stab1366",
    journal = "Mon. Not. Roy. Astron. Soc.",
    volume = "505",
    number = "2",
    pages = "2840--2869",
    year = "2021"
}

@article{Arnold:2021xtm,
    author = "Arnold, Christian and Li, Baojiu and Giblin, Benjamin and Harnois-D\'eraps, Joachim and Cai, Yan-Chuan",
    title = "{forge: the f(R)-gravity cosmic emulator project \textendash{} I. Introduction and matter power spectrum emulator}",
    eprint = "2109.04984",
    archivePrefix = "arXiv",
    primaryClass = "astro-ph.CO",
    doi = "10.1093/mnras/stac1091",
    journal = "Mon. Not. Roy. Astron. Soc.",
    volume = "515",
    number = "3",
    pages = "4161--4175",
    year = "2022"
}

@article{Kaushal:2021hqv,
    author = "Kaushal, Neerav and Villaescusa-Navarro, Francisco and Giusarma, Elena and Li, Yin and Hawry, Conner and Reyes, Mauricio",
    title = "{NECOLA: Toward a Universal Field-level Cosmological Emulator}",
    eprint = "2111.02441",
    archivePrefix = "arXiv",
    primaryClass = "astro-ph.CO",
    doi = "10.3847/1538-4357/ac5c4a",
    journal = "Astrophys. J.",
    volume = "930",
    number = "2",
    pages = "115",
    year = "2022"
}

@article{Jamieson:2022lqc,
    author = "Jamieson, Drew and Li, Yin and de Oliveira, Renan Alves and Villaescusa-Navarro, Francisco and Ho, Shirley and Spergel, David N.",
    title = "{Field-level Neural Network Emulator for Cosmological N-body Simulations}",
    eprint = "2206.04594",
    archivePrefix = "arXiv",
    primaryClass = "astro-ph.CO",
    doi = "10.3847/1538-4357/acdb6c",
    journal = "Astrophys. J.",
    volume = "952",
    number = "2",
    pages = "145",
    year = "2023"
}

@article{Moran:2022iwe,
    author = "Moran, Kelly R. and Heitmann, Katrin and Lawrence, Earl and Habib, Salman and Bingham, Derek and Upadhye, Amol and Kwan, Juliana and Higdon, David and Payne, Richard",
    title = "{The Mira\textendash{}Titan Universe \textendash{} IV. High-precision power spectrum emulation}",
    eprint = "2207.12345",
    archivePrefix = "arXiv",
    primaryClass = "astro-ph.CO",
    doi = "10.1093/mnras/stac3452",
    journal = "Mon. Not. Roy. Astron. Soc.",
    volume = "520",
    number = "3",
    pages = "3443--3458",
    year = "2023"
}

@article{Bose:2022vwi,
    author = "Bose, B. and Tsedrik, M. and Kennedy, J. and Lombriser, L. and Pourtsidou, A. and Taylor, A.",
    title = "{Fast and accurate predictions of the non-linear matter power spectrum for general models of Dark Energy and Modified Gravity}",
    eprint = "2210.01094",
    archivePrefix = "arXiv",
    primaryClass = "astro-ph.CO",
    doi = "10.1093/mnras/stac3783",
    journal = "Mon. Not. Roy. Astron. Soc.",
    volume = "519",
    number = "3",
    pages = "4780--4800",
    year = "2023"
}

@article{Bolliet:2023sst,
    author = "Bolliet, Boris and Spurio Mancini, Alessio and Hill, J. Colin and Madhavacheril, Mathew and Jense, Hidde T. and Calabrese, Erminia and Dunkley, Jo",
    title = "{High-accuracy emulators for observables in \ensuremath{\Lambda}CDM, Neff, \ensuremath{\Sigma}m\ensuremath{\nu}, and w cosmologies}",
    eprint = "2303.01591",
    archivePrefix = "arXiv",
    primaryClass = "astro-ph.CO",
    doi = "10.1093/mnras/stae1201",
    journal = "Mon. Not. Roy. Astron. Soc.",
    volume = "531",
    number = "1",
    pages = "1351--1370",
    year = "2024"
}

@article{Saez-Casares:2023olw,
    author = "S\'aez-Casares, I\~nigo and Rasera, Yann and Li, Baojiu",
    title = "{The e-MANTIS emulator: fast predictions of the non-linear matter power spectrum in $f(R)$CDM cosmology}",
    eprint = "2303.08899",
    archivePrefix = "arXiv",
    primaryClass = "astro-ph.CO",
    doi = "10.1093/mnras/stad3343",
    journal = "Mon. Not. Roy. Astron. Soc.",
    volume = "527",
    number = "3",
    pages = "7242--7262",
    year = "2024"
}

@article{Piras:2023aub,
    author = "Piras, D. and Spurio Mancini, A.",
    title = "{CosmoPower-JAX: high-dimensional Bayesian inference with differentiable cosmological emulators}",
    eprint = "2305.06347",
    archivePrefix = "arXiv",
    primaryClass = "astro-ph.CO",
    doi = "10.21105/astro.2305.06347",
    month = "5",
    year = "2023"
}

@article{Ho:2023alj,
    author = "Ho, Ming-Feng and Bird, Simeon and Fernandez, Martin A. and Shelton, Christian R.",
    title = "{MF-Box: multifidelity and multiscale emulation for the matter power spectrum}",
    eprint = "2306.03144",
    archivePrefix = "arXiv",
    primaryClass = "astro-ph.CO",
    doi = "10.1093/mnras/stad2901",
    journal = "Mon. Not. Roy. Astron. Soc.",
    volume = "526",
    number = "2",
    pages = "2903--2919",
    year = "2023"
}

@article{Bonici:2023xjk,
    author = "Bonici, Marco and Bianchini, Federico and Ruiz-Zapatero, Jaime",
    title = "{Capse.jl: efficient and auto-differentiable CMB power spectra emulation}",
    eprint = "2307.14339",
    archivePrefix = "arXiv",
    primaryClass = "astro-ph.CO",
    doi = "10.21105/astro.2307.14339",
    month = "7",
    year = "2023"
}

@article{Mauland:2023pjt,
    author = "Mauland, Renate and Winther, Hans A. and Ruan, Cheng-Zong",
    title = "{Sesame: A power spectrum emulator pipeline for beyond-\ensuremath{\Lambda}CDM models}",
    eprint = "2309.13295",
    archivePrefix = "arXiv",
    primaryClass = "astro-ph.CO",
    doi = "10.1051/0004-6361/202347892",
    journal = "Astron. Astrophys.",
    volume = "685",
    pages = "A156",
    year = "2024"
}

@article{Fiorini:2023fjl,
    author = "Fiorini, Bartolomeo and Koyama, Kazuya and Baker, Tessa",
    title = "{Fast production of cosmological emulators in modified gravity: the matter power spectrum}",
    eprint = "2310.05786",
    archivePrefix = "arXiv",
    primaryClass = "astro-ph.CO",
    doi = "10.1088/1475-7516/2023/12/045",
    journal = "JCAP",
    volume = "12",
    pages = "045",
    year = "2023"
}

@article{Bartlett:2023cyr,
    author = "Bartlett, Deaglan J. and Kammerer, Lukas and Kronberger, Gabriel and Desmond, Harry and Ferreira, Pedro G. and Wandelt, Benjamin D. and Burlacu, Bogdan and Alonso, David and Zennaro, Matteo",
    title = "{A precise symbolic emulator of the linear matter power spectrum}",
    eprint = "2311.15865",
    archivePrefix = "arXiv",
    primaryClass = "astro-ph.CO",
    doi = "10.1051/0004-6361/202348811",
    journal = "Astron. Astrophys.",
    volume = "686",
    pages = "A209",
    year = "2024"
}

@article{Trusov:2024mmw,
    author = "Trusov, Svyatoslav and Zarrouk, Pauline and Cole, Shaun",
    title = "{Neural network-based model of galaxy power spectrum: fast full-shape galaxy power spectrum analysis}",
    eprint = "2403.20093",
    archivePrefix = "arXiv",
    primaryClass = "astro-ph.CO",
    doi = "10.1093/mnras/staf285",
    journal = "Mon. Not. Roy. Astron. Soc.",
    volume = "538",
    number = "3",
    pages = "1789--1799",
    year = "2025"
}

@article{Bai:2024cgt,
    author = "Bai, Jiachen and Xia, Jun-Qing",
    title = "{FREmu: Power Spectrum Emulator for f(R) Gravity}",
    eprint = "2405.05840",
    archivePrefix = "arXiv",
    primaryClass = "astro-ph.CO",
    doi = "10.3847/1538-4357/ad55ef",
    journal = "Astrophys. J.",
    volume = "971",
    number = "1",
    pages = "11",
    year = "2024"
}

@article{Sui:2024wob,
    author = "Sui, Ce and Bartlett, Deaglan J. and Pandey, Shivam and Desmond, Harry and Ferreira, Pedro G. and Wandelt, Benjamin D.",
    title = "{syren-new: Precise formulae for the linear and nonlinear matter power spectra with massive neutrinos and dynamical dark energy}",
    eprint = "2410.14623",
    archivePrefix = "arXiv",
    primaryClass = "astro-ph.CO",
    doi = "10.1051/0004-6361/202452854",
    journal = "Astron. Astrophys.",
    volume = "698",
    pages = "A1",
    year = "2025"
}

@article{Kaushal:2025gwe,
    author = "Kaushal, Neerav and Giusarma, Elena and Reyes, Mauricio",
    title = "{nuGAN: Generative Adversarial Emulator for Cosmic Web with Neutrinos}",
    eprint = "2505.03936",
    archivePrefix = "arXiv",
    primaryClass = "astro-ph.CO",
    month = "5",
    year = "2025"
}

@article{SpurioMancini:2021ppk,
    author = "Spurio Mancini, Alessio and Piras, Davide and Alsing, Justin and Joachimi, Benjamin and Hobson, Michael P.",
    title = "{CosmoPower: emulating cosmological power spectra for accelerated Bayesian inference from next-generation surveys}",
    eprint = "2106.03846",
    archivePrefix = "arXiv",
    primaryClass = "astro-ph.CO",
    doi = "10.1093/mnras/stac064",
    journal = "Mon. Not. Roy. Astron. Soc.",
    volume = "511",
    number = "2",
    pages = "1771--1788",
    year = "2022"
}

@article{Jense:2024llt,
    author = "Jense, H. T. and Harrison, I. and Calabrese, E. and Spurio Mancini, A. and Bolliet, B. and Dunkley, J. and Hill, J. C.",
    title = "{A complete framework for cosmological emulation and inference with CosmoPower}",
    eprint = "2405.07903",
    archivePrefix = "arXiv",
    primaryClass = "astro-ph.CO",
    doi = "10.1093/rasti/rzaf002",
    journal = "RAS Tech. Instrum.",
    volume = "4",
    pages = "rzaf002",
    year = "2025"
}

@article{Gunther:2025xrq,
    author = {G\"unther, Sven and Balkenhol, Lennart and Fidler, Christian and Khalife, Ali Rida and Lesgourgues, Julien and Mosbech, Markus R. and Sharma, Ravi Kumar},
    title = "{OL\'E-- Online Learning Emulation in Cosmology}",
    eprint = "2503.13183",
    archivePrefix = "arXiv",
    primaryClass = "astro-ph.CO",
    reportNumber = "TTK-25-08",
    month = "3",
    year = "2025"
}

@article{SPT-3G:2022hvq,
    author = "Balkenhol, L. and others",
    collaboration = "SPT-3G",
    title = "{Measurement of the CMB temperature power spectrum and constraints on cosmology from the SPT-3G 2018 TT, TE, and EE dataset}",
    eprint = "2212.05642",
    archivePrefix = "arXiv",
    primaryClass = "astro-ph.CO",
    reportNumber = "FERMILAB-PUB-22-953-PPD",
    doi = "10.1103/PhysRevD.108.023510",
    journal = "Phys. Rev. D",
    volume = "108",
    number = "2",
    pages = "023510",
    year = "2023"
}

@article{SPT:2023jql,
    author = "Pan, Z. and others",
    collaboration = "SPT",
    title = "{Measurement of gravitational lensing of the cosmic microwave background using SPT-3G 2018 data}",
    eprint = "2308.11608",
    archivePrefix = "arXiv",
    primaryClass = "astro-ph.CO",
    doi = "10.1103/PhysRevD.108.122005",
    journal = "Phys. Rev. D",
    volume = "108",
    number = "12",
    pages = "122005",
    year = "2023"
}

@article{ACT:2020gnv,
    author = "Aiola, Simone and others",
    collaboration = "ACT",
    title = "{The Atacama Cosmology Telescope: DR4 Maps and Cosmological Parameters}",
    eprint = "2007.07288",
    archivePrefix = "arXiv",
    primaryClass = "astro-ph.CO",
    doi = "10.1088/1475-7516/2020/12/047",
    journal = "JCAP",
    volume = "12",
    pages = "047",
    year = "2020"
}

@article{ACT:2020frw,
    author = "Choi, Steve K. and others",
    collaboration = "ACT",
    title = "{The Atacama Cosmology Telescope: a measurement of the Cosmic Microwave Background power spectra at 98 and 150 GHz}",
    eprint = "2007.07289",
    archivePrefix = "arXiv",
    primaryClass = "astro-ph.CO",
    doi = "10.1088/1475-7516/2020/12/045",
    journal = "JCAP",
    volume = "12",
    pages = "045",
    year = "2020"
}

@article{ACT:2023kun,
    author = "Madhavacheril, Mathew S. and others",
    collaboration = "ACT",
    title = "{The Atacama Cosmology Telescope: DR6 Gravitational Lensing Map and Cosmological Parameters}",
    eprint = "2304.05203",
    archivePrefix = "arXiv",
    primaryClass = "astro-ph.CO",
    reportNumber = "FERMILAB-PUB-23-206-PPD",
    doi = "10.3847/1538-4357/acff5f",
    journal = "Astrophys. J.",
    volume = "962",
    number = "2",
    pages = "113",
    year = "2024"
}

@article{ACT:2023dou,
    author = "Qu, Frank J. and others",
    collaboration = "ACT",
    title = "{The Atacama Cosmology Telescope: A Measurement of the DR6 CMB Lensing Power Spectrum and Its Implications for Structure Growth}",
    eprint = "2304.05202",
    archivePrefix = "arXiv",
    primaryClass = "astro-ph.CO",
    reportNumber = "FERMILAB-PUB-23-237-PPD, FERMILAB-PUB-23-237-PPD",
    doi = "10.3847/1538-4357/acfe06",
    journal = "Astrophys. J.",
    volume = "962",
    number = "2",
    pages = "112",
    year = "2024"
}

@article{CMB-S4:2016ple,
    author = "Abazajian, Kevork N. and others",
    collaboration = "CMB-S4",
    title = "{CMB-S4 Science Book, First Edition}",
    eprint = "1610.02743",
    archivePrefix = "arXiv",
    primaryClass = "astro-ph.CO",
    reportNumber = "FERMILAB-FN-1024-A-AE",
    month = "10",
    year = "2016"
}

@article{SimonsObservatory:2018koc,
    author = "Ade, Peter and others",
    collaboration = "Simons Observatory",
    title = "{The Simons Observatory: Science goals and forecasts}",
    eprint = "1808.07445",
    archivePrefix = "arXiv",
    primaryClass = "astro-ph.CO",
    doi = "10.1088/1475-7516/2019/02/056",
    journal = "JCAP",
    volume = "02",
    pages = "056",
    year = "2019"
}

@article{Heymans:2020gsg,
    author = "Heymans, Catherine and others",
    title = "{KiDS-1000 Cosmology: Multi-probe weak gravitational lensing and spectroscopic galaxy clustering constraints}",
    eprint = "2007.15632",
    archivePrefix = "arXiv",
    primaryClass = "astro-ph.CO",
    doi = "10.1051/0004-6361/202039063",
    journal = "Astron. Astrophys.",
    volume = "646",
    pages = "A140",
    year = "2021"
}

@article{DES:2021wwk,
    author = "Abbott, T. M. C. and others",
    collaboration = "DES",
    title = "{Dark Energy Survey Year 3 results: Cosmological constraints from galaxy clustering and weak lensing}",
    eprint = "2105.13549",
    archivePrefix = "arXiv",
    primaryClass = "astro-ph.CO",
    reportNumber = "FERMILAB-PUB-21-221-AE, DES-2020-0617",
    doi = "10.1103/PhysRevD.105.023520",
    journal = "Phys. Rev. D",
    volume = "105",
    number = "2",
    pages = "023520",
    year = "2022"
}

@article{Eifler:2020vvg,
    author = "Eifler, Tim and others",
    title = "{Cosmology with the Roman Space Telescope \textendash{} multiprobe strategies}",
    eprint = "2004.05271",
    archivePrefix = "arXiv",
    primaryClass = "astro-ph.CO",
    doi = "10.1093/mnras/stab1762",
    journal = "Mon. Not. Roy. Astron. Soc.",
    volume = "507",
    number = "2",
    pages = "1746--1761",
    year = "2021"
}

@article{Saxena:2024rhu,
    author = "Saxena, Anchal and Meerburg, P. Daniel and Weniger, Christoph and Acedo, Eloy de Lera and Handley, Will",
    title = "{Simulation-based inference of the sky-averaged 21-cm signal from CD-EoR with REACH}",
    eprint = "2403.14618",
    archivePrefix = "arXiv",
    primaryClass = "astro-ph.CO",
    doi = "10.1093/rasti/rzae047",
    journal = "RAS Tech. Instrum.",
    volume = "3",
    number = "1",
    pages = "724--736",
    year = "2024"
}

@article{Floss:2023ylq,
    author = {Fl\"oss, Thomas and Meerburg, P. Daniel},
    title = "{Improving constraints on primordial non-Gaussianity using neural network based reconstruction}",
    eprint = "2305.07018",
    archivePrefix = "arXiv",
    primaryClass = "astro-ph.CO",
    doi = "10.1088/1475-7516/2024/02/031",
    journal = "JCAP",
    volume = "02",
    pages = "031",
    year = "2024"
}

@article{Chen:2006nt,
    author = "Chen, Xingang and Huang, Min-xin and Kachru, Shamit and Shiu, Gary",
    title = "{Observational signatures and non-Gaussianities of general single field inflation}",
    eprint = "hep-th/0605045",
    archivePrefix = "arXiv",
    reportNumber = "SLAC-PUB-11840, MAD-TH-06-3, UFIFT-HEP-06-9, SU-ITP-06-12, CU-TP-1147",
    doi = "10.1088/1475-7516/2007/01/002",
    journal = "JCAP",
    volume = "01",
    pages = "002",
    year = "2007"
}

@article{Jung:2025nss,
    author = "Jung, Gabriel and Citran, Michele and van Tent, Bartjan and Dumilly, L\'ea and Aghanim, Nabila",
    title = "{Constraints on primordial non-Gaussianity from Planck PR4 data}",
    eprint = "2504.00884",
    archivePrefix = "arXiv",
    primaryClass = "astro-ph.CO",
    month = "4",
    year = "2025"
}

@article{Karagiannis:2018jdt,
    author = "Karagiannis, Dionysios and Lazanu, Andrei and Liguori, Michele and Raccanelli, Alvise and Bartolo, Nicola and Verde, Licia",
    title = "{Constraining primordial non-Gaussianity with bispectrum and power spectrum from upcoming optical and radio surveys}",
    eprint = "1801.09280",
    archivePrefix = "arXiv",
    primaryClass = "astro-ph.CO",
    doi = "10.1093/mnras/sty1029",
    journal = "Mon. Not. Roy. Astron. Soc.",
    volume = "478",
    number = "1",
    pages = "1341--1376",
    year = "2018"
}

@article{DESI:2025ejh,
    author = "Elbers, W. and others",
    collaboration = "DESI",
    title = "{Constraints on Neutrino Physics from DESI DR2 BAO and DR1 Full Shape}",
    eprint = "2503.14744",
    archivePrefix = "arXiv",
    primaryClass = "astro-ph.CO",
    reportNumber = "FERMILAB-PUB-25-0168-PPD",
    month = "3",
    year = "2025"
}

\end{document}